\documentclass[12pt,a4paper]{article}

\usepackage{epsfig}
\usepackage{amsmath,amsfonts,amssymb}
\usepackage{cite}
\usepackage{colortbl}
\usepackage{pifont}

\providecommand{\openone}{\leavevmode\hbox{\small1\kern-3.8pt\normalsize1}}

\begin{document}

\vspace*{-3cm}
\begin{flushright}
CERN-TH-2017-013
\end{flushright}
\vspace{0.5cm}

\begin{center}
\begin{Large}
{\bf Single top polarisation as a window to \\[1mm] new physics}
\end{Large}

\vspace{0.5cm}
J.~A.~Aguilar--Saavedra$^{a}$, C.~Degrande$^{b}$, S.~Khatibi$^c$ \\[1mm]
\begin{small}
{$^a$ Departamento de F\'{\i}sica Te\'orica y del Cosmos, 
Universidad de Granada, \\ E-18071 Granada, Spain} \\ 
{$^b$ CERN, Theory Department, Geneva 23 CH-1211, Switzerland} \\
{$^c$ School of Particles and Accelerators, Institute for Research in Fundamental Sciences (IPM) P.O. Box 19395-5531, Tehran, Iran } \\
 
\end{small}
\end{center}

\begin{abstract}
We discuss the effect of heavy new physics, parameterised in terms of four-fermion operators, in the polarisation of single top (anti-)quarks in the $t$-channel process at the LHC. It is found that for operators involving a right-handed top quark field the relative effect on the longitudinal polarisation is twice larger than the relative effect on the total cross section. This enhanced dependence on possible four-fermion contributions makes the polarisation measurements specially interesting, in particular at high momenta. 
\end{abstract}

Single top quark production is the sub-dominant source of top quarks in high-energy collisions at the Large Hadron Collider (LHC). The three production modes, $t$-channel, $s$-channel and $tW$ associated production, are sensitive to various manifestations of new physics beyond the Standard Model (SM), such as new particles exchanged in the $s$ or $t$ channels~\cite{Tait:2000sh,Boos:2006xe,Arhrib:2006sg,Drueke:2014pla}, a modified $tbW$ interaction~\cite{Boos:1999dd,Chen:2005vr,AguilarSaavedra:2008gt,Zhang:2010dr}, top flavour-changing neutral currents~\cite{Hosch:1997gz,Han:1998tp,Kidonakis:2003sc,Ferreira:2005dr,Ferreira:2006xe,AguilarSaavedra:2010rx,Etesami:2010ig} or effective four-fermion interactions~\cite{Zhang:2010dr,AguilarSaavedra:2010zi}. Most of the phenomenological studies addressing the effect of new physics in single top production have concentrated on the total cross section, whereas the top quark polarisation, which can be measured with similar or better precision, deserves a more detailed investigation. 

In the $t$-channel process, single top quarks are produced with a large longitudinal polarisation $P_z \simeq 0.9$ in the direction $\hat z$ of the spectator jet in the top quark rest frame~\cite{Mahlon:1999gz}, and a slightly smaller polarisation $P_z \simeq 0.8$ for anti-quarks. (In the $t$-channel process $qg \to q' t \bar b$ the spectator jet $j$ is the light quark $q'$.) The top quark polarisation can also be measured along two additional orthogonal axes, the so-called ``transverse'' ($\hat x$, within the production plane) and ``normal'' ($\hat y$, orthogonal to it). The definitions of the three axes are~\cite{Aguilar-Saavedra:2014eqa}
\begin{equation}
\hat z = \frac{\vec p_j}{|\vec p_j|} \,,\quad
\hat y = \frac{\vec p_j \times \vec p_q}{|\vec p_j \times \vec p_q|} \,,\quad
\hat x = \hat y \times \hat z \,,
\label{ec:axes}
\end{equation}
with the direction of the initial quark $q$ taken as the direction of the proton closest to the spectator jet. All momenta are taken in the top rest frame. In the SM, $P_x \simeq 0$ for quarks and $P_x \simeq 0.1$ for anti-quarks at the tree level, and in both cases $P_y = 0$ because a normal polarisation requires a non-trivial complex phase in the amplitude. The effect in the top polarisation of new physics at the loop level, which manifests in the form of anomalous $tbW$ couplings, has been studied before~\cite{Espriu:2002wx,Aguilar-Saavedra:2014eqa}. In this Letter we will fill the gap by studying the effect in the top polarisation of new heavy resonances, described in terms of four-fermion effective interactions.\footnote{On-shell particles exchanged in the $s$ channel obviously produce large effects in the top polarisation as well~\cite{Gopalakrishna:2010xm}, if they are within kinematical reach.} 

We work in a framework of the SM extended with an effective Lagrangian containing dimension-six gauge invariant effective operators $O_i$,
\begin{equation}
\mathcal{L}_\text{eff} =  \sum_i \frac{C_i}{\Lambda^2} \, O_i + \text{h.c.} \,,
\end{equation}
where $\Lambda$ is the new physics scale and $C_i$ are dimensionless coefficients. 
Conventionally, we will add the Hermitian conjugate even if the operators $O_i$ are Hermitian. As aforementioned, in this work we restrict ourselves to operators involving four quark fields. There is a huge number (185) of such operators contributing to single top production~\cite{AguilarSaavedra:2010zi}, so we need some guiding principle to select among the possible four-fermion contributions and still obtain results that are representative of the phenomenology induced by such operators.

\begin{figure}[htb]
\begin{center}
\begin{tabular}{ccc}
\includegraphics[height=3.5cm,clip=]{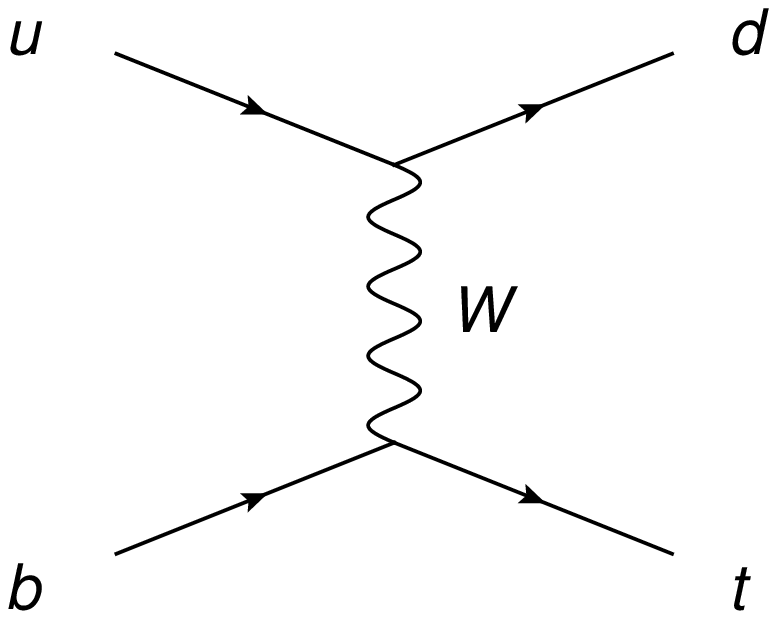} & \quad\quad\quad &
\includegraphics[height=3.5cm,clip=]{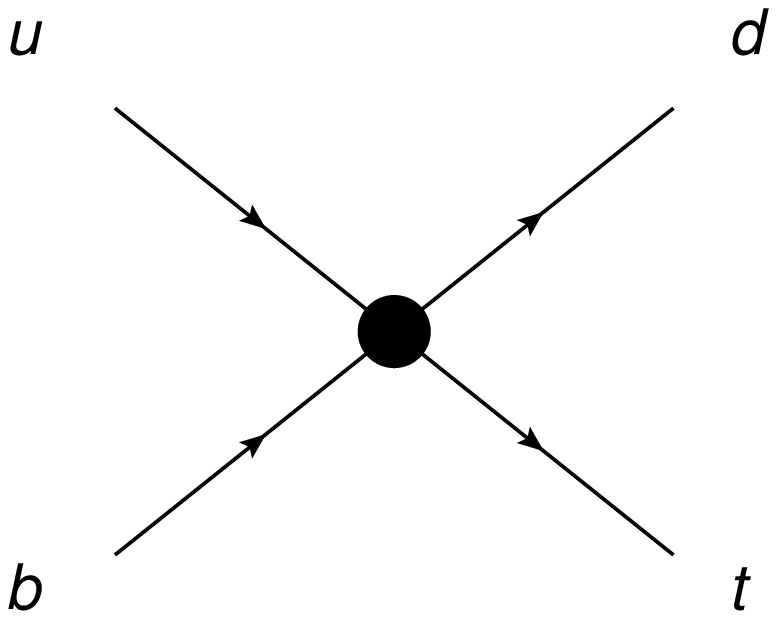} 
\end{tabular}
\end{center}
\caption{Left: sample $2 \to 2$ diagram for $t$-channel single top production. Right: four-fermion contribution.}
\label{fig:1}
\end{figure}

Neglecting the charged-current mixing, the amplitudes for the leading SM $t$-channel single top processes involve the exchange of a $W$ boson between two fermion lines, one with $u$ and $d$ quarks and the second one with $b$ and $t$ quarks, such as the $2 \to 2$ diagram depicted in Fig.~\ref{fig:1} (left), plus diagrams related by crossing symmetry.
There are sixteen independent four-fermion operators with these quark fields~\cite{AguilarSaavedra:2010zi}, depicted symbolically in the right panel. All of them can contribute to these leading single top production processes. But not all of them are expected to give qualitatively different effects on the cross section and polarisation. There are six independent operators that involve the same colour flow as the SM $u b \to d t$ amplitude, namely the diagram in Fig.~\ref{fig:1} (left), which can be taken as
\begin{align}
& O_{qq}^{(3,3311)} = (\bar q_{L3} \gamma^\mu \tau^I q_{L3}) (\bar q_{L1} \gamma_\mu \tau^I q_{L1}) 
\supset 2 (\bar t_L \gamma^\mu b_L) (\bar d_L \gamma_\mu u_L)   \,, \notag \\
& O_{du}^{(1133)} = (\bar d_R \gamma^\mu u_R) (\bar t_R \gamma_\mu b_R) \,, \notag \\
& O_{qu}^{(1133)} = (\bar q_{L1} u_R) (\bar t_R q_{L3}) \supset (\bar d_L u_R) (\bar t_R b_L) \,, \notag \\
& O_{qd}^{(3311)} = (\bar q_{L3} b_R) (\bar d_R q_{L1}) \supset (\bar t_L b_R) (\bar d_R u_L) \,, \notag \\
& O_{qud}^{(1,3311)} = (\bar q_{L3} t_R) \epsilon (\bar q_{L1} d_R)^T \supset - (\bar b_L t_R) (\bar u_L d_R) \,, \notag \\
& O_{qud}^{(1,1133)} = (\bar q_{L1} u_R) \epsilon (\bar q_{L3} b_R)^T \supset - (\bar d_L u_R) (\bar t_L b_R) \,.
\label{ec:op}
\end{align}
We use standard notation with $q_{Li} = (u_{Li} d_{Li})^T$ the left-handed doublet of the $i^\text{th}$ generation, $\tau^I$ the Pauli matrices, and $\epsilon = i \tau^2$. Six additional operators
\begin{align}
& O_{qq}^{(1,3113)} = (\bar q_{L3} \gamma^\mu q_{L1}) (\bar q_{L1} \gamma_\mu q_{L3}) 
\supset 2 (\bar t_L \gamma^\mu u_{L}) (\bar d_{L} \gamma_\mu b_{L})   \,, \notag \\
& O_{ud}^{(3113)} = (\bar t_R \gamma^\mu u_R) (\bar d_R \gamma_\mu b_R) \,, \notag \\
& O_{uq}^{(3113)} = (\bar t_R \gamma^\mu u_R) (\bar q_{L1} \gamma_\mu q_{L3}) \supset  (\bar t_R \gamma^\mu u_R) (\bar d_{L} \gamma_\mu b_{L})\,, \notag \\
& O_{dq}^{(1331)} = (\bar d_R \gamma^\mu b_R) (\bar q_{L3} \gamma_\mu q_{L1}) \supset
 (\bar d_R \gamma^\mu b_R) (\bar t_{L} \gamma_\mu u_{L})   \,, \notag \\
& O_{qud}^{(1,1331)} = (\bar q_{L1} t_R) \epsilon (\bar q_{L3} d_R)^T \supset (\bar u_L t_R) (\bar b_L d_R) \,, \notag \\
& O_{qud}^{(1,3113)} = (\bar q_{L3} u_R) \epsilon (\bar q_{L1} b_R)^T \supset (\bar t_L u_R) (\bar d_L b_R) \,,
\end{align}
give similar terms but with the colour indices contracted between $t,u$ and $d,b$, respectively. For the operators with two or more right-handed fields this difference has no effect in the colour-averaged differential cross sections at the partonic level, because the diagrams including four-fermion operators do not interfere with the SM amplitudes for massless $u$ and $d$ quarks. On the other hand, for the two operators with four left-handed fields, the interference of $O_{qq}^{(1,3113)}$ is damped by a colour factor of $1/3$ while for $O_{qq}^{(3,3311)}$ it is not; however, neither of these operators change the top polarisation significantly, therefore the former operator does not lead to a qualitatively different phenomenology either. The remaining four operators
\begin{align}
& O_{qud}^{(8,3311)} = (\bar q_{L3}  \lambda^a t_R) \epsilon (\bar q_{L1} \lambda^a d_R)^T \supset - (\bar b_L \lambda^a t_R) (\bar u_L \lambda^a d_R) \,, \notag \\
& O_{qud}^{(8,1133)} = (\bar q_{L1} \lambda^a u_R) \epsilon (\bar q_{L3} \lambda^a b_R)^T \supset - (\bar d_L \lambda^a u_R) (\bar t_L \lambda^a b_R) \,, \notag \\
& O_{qud}^{(8,1331)} = (\bar q_{L1} \lambda^a t_R) \epsilon (\bar q_{L3} \lambda^a d_R)^T \supset (\bar u_L \lambda^a t_R) (\bar b_L \lambda^a d_R) \,, \notag \\
& O_{qud}^{(8,3113)} = (\bar q_{L3} \lambda^a u_R) \epsilon (\bar q_{L1} \lambda^a b_R)^T \supset (\bar t_L \lambda^a u_R) (\bar d_L \lambda^a b_R) 
\end{align}
are also expected to yield similar results up to rescaling factors, because they do not interfere with the SM amplitude.
 
We do not consider here the 80 additional four-fermion operators obtained by replacing $u \to c$ and/or $d \to s,b$ in the former set. Either their effect is smaller than the ones considered, because of parton distribution functions (PDFs), if the second generation or $b$ quarks are in the initial state, or the effect in the cross section and polarisation is similar, if they are in the final state. By the same reckoning, the effect of flavour-changing neutral four-fermion operators (involving three charge-$2/3$ light quarks plus a top quark) is expected to be quite similar. Therefore, we restrict our study to the six operators in (\ref{ec:op}), which generate the effective Lagrangian
\begin{eqnarray}
\mathcal{L}_\text{eff} & = & \frac{1}{\Lambda^2} \left[
C_{qq} \, (\bar t_L \gamma^\mu b_L) (\bar d_L \gamma_\mu u_L) 
+ C_{du} (\bar t_R \gamma^\mu b_R) (\bar d_R \gamma_\mu u_R) 
+ C_{qu}  (\bar t_R b_L)  (\bar d_L u_R) \right. \notag \\
& & \left. + C_{qd} (\bar t_L b_R) (\bar d_R u_L) 
- C_{qud_R}  (\bar t_R b_L) (\bar d_R u_L)
- C_{qud_L}  (\bar t_L b_R) (\bar d_L u_R)  \right] + \text{h.c.}
\label{ec:lagrsimpl}
\end{eqnarray}
In order to ease the notation, we have dropped flavour superindices in the effective operator coefficients and introduced a chirality label to distinguish the coefficients of the two $O_{qud}$ operators,
$C_{qud_R} = C_{qud}^{(1,3311)\,*}$, $C_{qud_L} = C_{qud}^{(1,1133)}$. A factor of 4 has been absorbed in the definition of $C_{qq}$.

This Lagrangian is implemented in {\scshape Feynrules}~\cite{Alloul:2013bka} in order to perform our calculations with {\scshape MadGraph5\_aMC@NLO}~\cite{Alwall:2014hca} using the Universal Feynrules Output~\cite{Degrande:2011ua}.
We use a centre-of-mass energy of 13 TeV. Our results include $ug \to t \bar b d$, $\bar d g \to t \bar b \bar u$ and also $u \bar d \to t \bar b g$, which is $s$-channel production with the radiation of an extra gluon, and the charge conjugate processes for top anti-quark production. The latter process has a collinear enhancement for low transverse momentum ($p_T$) of the spectator jet, in this case the gluon; Since experimental measurements require the presence of the spectator jet we set a lower cut $p_T \geq 20$ GeV. We use the {\scshape CTEQ6L1} PDFs~\cite{Nadolsky:2008zw} and the default factorisation and renormalisation scales in the generator.\footnote{The default scale of {\scshape MadGraph5\_aMC@NLO} combines the diagram enhancement method \cite{Maltoni:2002qb} and the CKKW clustering \cite{Catani:2001cc} in order to have a dynamical scale that depends on the topology of the event generated. For the final state radiation diagram, the scale will be chosen as the invariant mass, while in presence of initial state radiation the scale will be chosen as the average of the transverse mass of the last particles of the clustering.}
Processes with initial $s$, $c$ quarks are included as well but without four-fermion operator contributions, which as seen in Eq.~(\ref{ec:lagrsimpl}) only involve $u$, $d$, $t$ and $b$ fields.
We compute in {\scshape MadGraph5\_aMC@NLO} the processes including the top decay $t \to W b \to \ell \nu b$. The produced events are then analysed to extract the fraction of events for which the angle $\theta$ between the lepton momentum in the top rest frame and the $\hat z$ direction is larger or smaller than $\pi/2$. This asymmetry is directly related to $P_z$~\cite{Kuhn:1981md},
\begin{equation}
P_z = 2 \frac{\sigma(\cos \theta > 0) - \sigma(\cos \theta < 0)}{\sigma} \,.
\label{ec:Pcos}
\end{equation}
Our calculations are done at leading order (LO). The longitudinal polarisation obtained using the $2 \to 3$ process is known to agree very well with the full next-to-leading order (NLO) calculation~\cite{Schwienhorst:2010je}, for example at 7 TeV $P_z = 0.90$ for top quarks and $P_z = -0.86$ for antiquarks at LO, while at NLO $P_z = 0.91$ and $P_z = -0.86$, respectively.

\begin{figure}[htb]
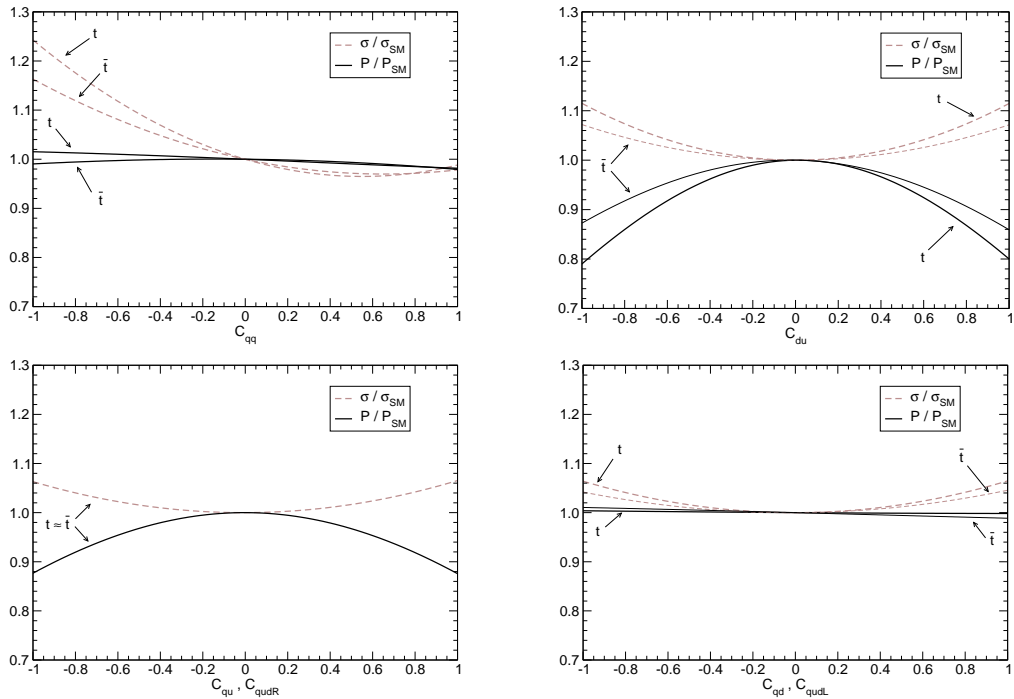

\begin{center}
\begin{tabular}{ccc}
\includegraphics[height=4.5cm,clip=]{Oqq.eps} & \quad\quad &
\includegraphics[height=4.5cm,clip=]{Odu.eps} \\
\includegraphics[height=4.5cm,clip=]{Oqu.eps} & \quad\quad &
\includegraphics[height=4.5cm,clip=]{Oqd.eps} \\
\end{tabular}
\end{center}
\caption{Dependence of the cross section and polarisation on effective operator coefficients, normalised to their SM values, for $\Lambda = 1$ TeV.}
\label{fig:depC}
\end{figure}

We only study the effect of one effective operator at a time, because their interference vanishes in most cases, or is suppressed by the $b$ quark mass otherwise. Therefore, the simultaneous inclusion of two or more operators does not lead to qualitatively new effects. The operator coefficients are taken real. ($C_{qq}$ is always real due to the Hermiticity of the effective operator.) We have not found significant variations in the transverse polarisation $P_x$ nor, for complex operator coefficients, in the normal polarisation $P_y$. The dependence of the total cross section and longitudinal polarisation on the effective operator coefficients is given by
\begin{eqnarray}
\sigma & = & A_0 + A_\text{int} \, \frac{C}{\Lambda^2} + A_2 \left( \frac{C}{\Lambda^2} \right)^2 \,, \notag \\
P_z & = &  \left[ B_0 + B_\text{int} \, \frac{C}{\Lambda^2} + B_2  \left( \frac{C}{\Lambda^2} \right)^2 \right] /\sigma \,,
\label{ec:sP}
\end{eqnarray}
with $A_0 = 93.7$ pb ($51.6$ pb) the SM cross section for top quarks (anti-quarks) and $B_0 = 81.6$ pb ($41.8$ pb). The values of the remaining coefficients are collected in Table~\ref{tab:ABt}. They are extracted from quadratic fits to $\sigma$ and $\sigma P_z$, numerically obtained with the Monte Carlo calculations for several values of each coefficient.
The dependence of the cross section and polarisation is depicted in Fig.~\ref{fig:depC}. For the operators involving right-handed fields $t_R$, it is seen that the effect in the top quark polarisation is twice larger than in the total cross section, the reason being that for these operators $A_2 \sim - B_2$, while $A_0 \sim B_0$ (the SM polarisation $P_z = B_0 / A_0$ is close to unity). The relation $A_2 \sim - B_2$ can be understood because these contributions correspond to the pure four-fermion terms, and those involving $t_R$ tend to produce positive helicity tops, contributing with negative sign to the numerator of (\ref{ec:Pcos}) and with positive sign to the denominator. Of course, for massive particles the helicity and chirality eigenstates do not coincide, hence the relation is expected to be only approximate. 

\begin{table}[htb]
\begin{center}
\begin{tabular}{lccccc|ccccc}
& \multicolumn{4}{c}{$t$} && \multicolumn{4}{c}{$\bar t$} \\
& $B_\text{int}$ & $B_2$ & $A_\text{int}$ & $A_2$ &&& $B_\text{int}$ & $B_2$ & $A_\text{int}$ & $A_2$ \\ 
\hline
$C_{qq}$                      & $-11.8$ & $9.3$  & $-11.9$ & $10.8$ &&
                                     & $-4.1$  & $2.3$   & $-4.8$   & $3.7$  \\
$C_{du}$                      &  $0$     & $-9.2$  & $0$       & $10.8$ &&
                                     & $0$      & $-3.0$  & $0$       & $3.7$ \\
$C_{qu}$, $C_{qudR}$ &        $0$ & $-5.4$ &        $0$ & $6.0$ &&
                                     &         $0$ & $-2.1$ & $0$ & $2.3$ \\
$C_{qd}$, $C_{qudL}$ &        $0$ & $5.3$ &        $0$ & $6.0$ &&
                                     &        $0$ & $1.8$ & $0$ & $2.3$ \\
\end{tabular}
\end{center}
\caption{Numerical values of the coefficients in Eqs.~(\ref{ec:sP}) for the different four-fermion operators. $A_\text{int}$, $B_\text{int}$ are given in units of pb TeV$^2$, and $A_2$, $B_2$ in units of pb TeV$^4$.
\label{tab:ABt}}
\end{table}

\begin{figure}[tb]
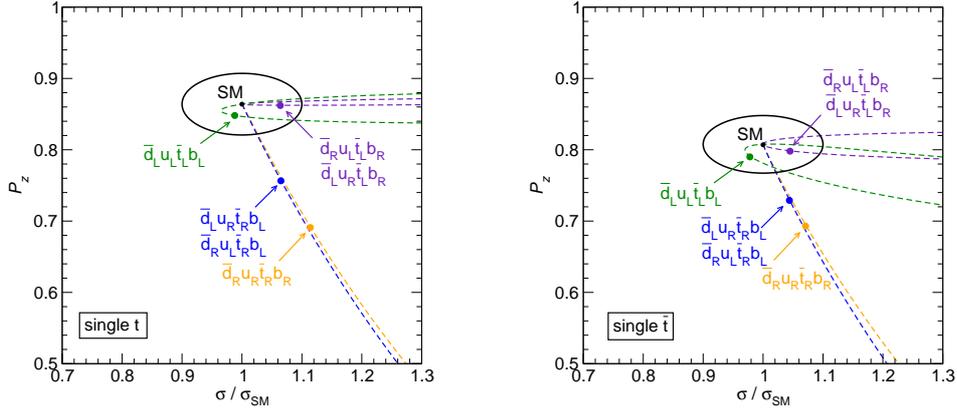

\begin{center}
\begin{tabular}{ccc}
\includegraphics[height=5.4cm,clip=]{PzS-t.eps} & \quad \quad
\includegraphics[height=5.4cm,clip=]{PzS-tbar.eps} 
\end{tabular}
\end{center}
\caption{Effect of four-fermion contributions in the single top cross section (normalised to the SM value) and polarisation, for top quarks (left) and anti-quarks (right). The black dots and ellipses represent the SM predictions and expected uncertainties. The points corresponding to operator coefficients $C / \Lambda^2 = 1~\text{TeV}^{-2}$ are indicated.}
\label{fig:summ}
\end{figure}

The relative effect of four-fermion contributions on the single top cross section and polarisation is compactly presented in Fig.~\ref{fig:summ}. As for the total cross section we take a 10\% relative error, assuming some improvement over current measurements at 13 TeV, $\sigma_t = 156 \pm 5\;\text{(stat)} \pm 27 \;\text{(syst)}$ pb, $\sigma_{\bar t} = 91 \pm 4 \;\text{(stat)} \pm 18 \;\text{(syst)}$ pb~\cite{Aaboud:2016ymp}, $\sigma_{t+\bar t} = 232 \pm 13\;\text{(stat)} \pm 28 \;\text{(syst)}$ pb~\cite{Sirunyan:2016cdg}.
Polarisation measurements have not yet been performed with Run 2 data, but for the full Run 1 data set we have $P_z = 0.97 \pm 0.05\;\text{(stat)} \pm 0.11 \;\text{(syst)}$~\cite{Aaboud:2017aqp},
$P_z = 0.52 \pm 0.06\;\text{(stat)} \pm 0.20 \;\text{(syst)}$~\cite{Khachatryan:2015dzz}.
Taking into account the fact that part of the systematic uncertainty in the polarisation measurement results from the size of Monte Carlo samples, we will assume an improvement to a 5\% uncertainty in its measurement. Under these conditions, it is clear that $P_z$ is a far more sensitive probe of the presence of four-fermion contributions involving right-handed top fields than the total cross section: not only the variation in $P_z$ is larger but also the precision in its measurement is better.

As it is well known, the effect of four-fermion interactions at high energies is enhanced, because these non-renormalisable terms do not include a propagator as the SM contributions do. Setting for instance a lower cut $p_T \geq 200$ GeV on the top quark transverse momentum, the SM contributions are reduced by an order of magnitude, $A_0 = 4.0$ pb, $B_0 = 2.7$ pb for top quarks, and $A_0 = 1.8$, $B_0 = 1.2$ for anti-quarks. The interference with four-fermion operators is similarly suppressed, while the quadratic four fermion contributions are of the same order, see Table~\ref{tab:ABtHI}. Therefore, the sensitivity to the presence of these contributions is greatly enhanced. For example, for $C_{du}/\Lambda^2 = 0.2~\text{TeV}^{-2}$, the variation in $P_z$ is as large as 20\% of its value. These deviations are of particular interest: in contrast with the differential cross section, polarisation measurements are expected to be rather insensitive to the uncertainties on the PDFs at the tails of the distributions, although the measurement for very boosted tops may involve other uncertainties. An estimate of the precision in high-$p_T$ polarisation measurements is difficult, and we look forward to actual measurements by the ATLAS and CMS Collaborations.

\begin{table}[htb]
\begin{center}
\begin{tabular}{lccccc|ccccc}
& \multicolumn{4}{c}{$t$} && \multicolumn{4}{c}{$\bar t$} \\
& $B_\text{int}$ & $B_2$ & $A_\text{int}$ & $A_2$ &&& $B_\text{int}$ & $B_2$ & $A_\text{int}$ & $A_2$ \\ 
\hline
$C_{qq}$                      & $-0.6$    & $7.1$    & $-0.7$    & $8.2$ &&
                                     & $\sim 0$ & $2.2$    & $\sim 0$ & $2.6$ \\
$C_{du}$                      &        $0$ & $-7.2$   &        $0$ & $8.2$  &&
                                     &        $0$ & $-2.4$   &        $0$ & $2.6$\\
$C_{qu}$, $C_{qudR}$ &        $0$ & $-4.6$  &        $0$ & $4.8$  &&
                                      &       $0$ & $-1.5$  &       $0$ & $1.8$    \\
$C_{qd}$, $C_{qudL}$ &        $0$ & $4.4$    &        $0$ & $4.8$  &&
                                     &        $0$ & $1.6$   &        $0$ &  $1.8$ \\
\end{tabular}
\end{center}
\caption{Numerical values of the coefficients in Eqs.~(\ref{ec:sP}) for the different four-fermion operators, with a lower cut $p_T \geq 200$ GeV for the top quark. $A_\text{int}$, $B_\text{int}$ are given in units of pb TeV$^2$, and $A_2$, $B_2$ in units of pb TeV$^4$.}
\label{tab:ABtHI}
\end{table}

To summarise, in this Letter we have addressed the influence of four-fermion contributions in the cross section and polarisation of top quarks in the $t$-channel process at the LHC. The possible effect of operators involving $t_R$ in the longitudinal polarisation is significant, especially at high energies. Our results highlight the importance of polarisation measurements, and especially motivate further measurements at high transverse momenta. Such measurements will be feasible with the high statistics that will be achieved at the LHC Run 2.

\section*{Acknowledgements}
S.K. thanks the CERN Theory Department and the University of Granada for their hospitality during the realisation of this work.
This research has been supported by MINECO Projects  FPA 2016-78220-C3-1-P, FPA 2013-47836-C3-2-P (including ERDF), Junta de Andaluc\'{\i}a Project FQM-101, Iran National Science Foundation (INSF) and  European Commission through the contract PITN-GA-2012-316704 (HIGGSTOOLS).

\end{document}